\documentclass[aps, twocolumn, superscriptaddress, floatfix, letterpaper,
preprintnumbers]{revtex4}
\usepackage{graphicx}

\newcommand{\Dirac}{\rlap {\hspace{-0.5mm} \slash} D}
\newcommand{\be}{\begin{eqnarray}}
\newcommand{\ee}{\end{eqnarray}}

\newcommand{\mat}{\left ( \begin{array}{cc}}
\newcommand{\emat}{\end{array} \right )}
\newcommand{\matf}{\left ( \begin{array}{cccc}}
\newcommand{\ematf}{\end{array} \right )}
\newcommand{\matt}{\left \begin{array}{ccc}}
\newcommand{\ematt}{\end{array} \right )}
\newcommand{\vect}{\left ( \begin{array}{c}}
\newcommand{\evect}{\end{array} \right )}
\newcommand{\Tr}{\rm Tr}
\newcommand{\nn}{\nonumber }

\begin{document}

\title{Isospin Chemical Potential and the QCD
  Phase Diagram at Nonzero Temperature and Baryon Chemical Potential}
\author{D.~Toublan and J.~B.~Kogut}
\affiliation{Department of Physics, University of Illinois at
Urbana-Champaign,
Urbana, IL 61801, USA}

\date{January 21, 2003}

\begin{abstract}
We use the Nambu--Jona-Lasinio model to study the effects of the isospin
chemical potential on the QCD phase diagram at nonzero temperature and
baryon chemical potential. We find that the phase diagram is
qualitatively altered by a small isospin chemical
potential. There are {\it two} first order phase transitions that end
in {\it two} critical endpoints, and there are {\it two} crossovers at
low baryon chemical potential. These results have important consequences
for systems where {\it both} baryon and isospin
chemical potentials are nonzero, such as heavy ion collision experiments.
Our results are in complete agreement with those recently obtained in
a Random Matrix Model.
\end{abstract}


\maketitle

\renewcommand{\theequation}{\arabic{equation}}
\setcounter{equation}{0} {\it Introduction.} -- QCD at nonzero
temperature and baryon chemical potential plays a fundamental role
in many different physical systems. Two important ones are neutron
stars, which probe the low temperature and high baryon chemical
potential domain, and heavy ion collision experiments, which probe
the high temperature and low baryon chemical potential domain. In
the last few years, various models have been used to predict the
main characteristics of the QCD phase diagram at nonzero
temperature and baryon chemical potential \cite{RWreview}. The
existence of color superconducting phases at low temperature and
high baryon chemical potential \cite{ARW,RSSV}, as well as the
presence of a tricritical point at intermediate temperature and
baryon chemical potential \cite{RMTphaseD,BR} are among the most
important results. These features have numerous phenomenological
consequences. Some of the effects of a nonzero isospin chemical
potential have also been studied, but only in the low temperature
and high baryon chemical potential domain 
\cite{Bedaque,ABRloff,BO,SPL,NBO,SRP}.
However systems such as heavy ion collision experiments that
explore the high temperature and low baryon chemical potential
domain also have a nonzero isospin chemical potential. Therefore,
there is a clear need to study the effects of a nonzero isospin
chemical potential on the whole QCD phase diagram at nonzero
temperature and baryon chemical potential.

Most of our knowledge of the QCD phase diagram at nonzero
temperature $T$ is restricted to either zero baryon chemical
potential $\mu_B$, or to zero isospin chemical potential $\mu_I$.
At $\mu_B$=$\mu_I$=$0$ numerical lattice simulations and effective
theories predict that the ground state corresponds to a hadronic
phase at low temperature and to a quark-gluon-plasma phase at high
temperature. For nonzero quark masses, there is no order parameter
that distinguishes between these two phases. For QCD with two
flavors, a crossover is expected at $T\sim170$ MeV \cite{LATTrev}.
This crossover extends into the phase diagram at nonzero baryon
and isospin chemical potentials.

Numerical lattice simulations have explored the high temperature
and small baryon chemical potential domain
\cite{FK,Owe,HK,Maria,Crompton,KS}. It was found that these high
temperature crossover lines at $\mu_B\ll\Lambda_{ \rm QCD}$ and
$\mu_I=0$, and at $\mu_B=0$ and $\mu_I$ much smaller than the pion
mass are identical \cite{HK,KS}.

At zero baryon chemical potential, both effective theories and
numerical lattice simulations predict the existence of a
superfluid pion condensation phase for high enough $\mu_I$
\cite{SS,KS,TV,KST-KSTVZ,KTS,STV,KTV}. At zero temperature a
second order phase transition at a critical isospin chemical
potential, $\mu_I^{\rm crit}$, equal to half the pion mass
separates the hadronic phase from the pion condensation phase.
When the temperature is increased, this second order phase
transition line ends in a tricritical point and the phase
transition becomes first order \cite{KTS,STV}.

At nonzero baryon chemical potential, standard numerical lattice
simulations do not work. Therefore our knowledge of the QCD phase
diagram at nonzero $T$ and $\mu_B$ relies
exclusively on effective
theories, such as Nambu--Jona-Lasinio and Random Matrix models.
At temperatures smaller than a few tens of MeV, an increase in $\mu_B$
leads to a crystalline LOFF
phase \cite{ABRloff}. If $\mu_B$ is further increased, the ground
state corresponds to a color superconductor. Both of these phase
transitions are of first order. If the temperature is increased to a
few tens of MeV the LOFF and color superconducting phases disappear,
and a
first order phase transition directly separates the
hadronic phase from the quark-gluon-plasma phase. At zero
quark mass, this first
order line ends in a tricritical endpoint at $T\sim 100{\rm MeV}$,
where a second order phase transition starts \cite{RMTphaseD,BR}.

Finally, the only available theoretical study of the {\it whole} phase
diagram at nonzero $T$, $\mu_B$, and $\mu_I$ has been performed
using a Random Matrix model \cite{KTV}. It was found that a small
isospin chemical potential induces two first order phase
transitions at low $T$ that end in two critical endpoints, and
there are two crossovers at low $\mu_B$. Because of the
phenomenological implications of these results, it is essential to
try to reproduce them within other models.

In this letter, we study the QCD phase diagram at nonzero $T$,
$\mu_B$, and $\mu_I$  for two quark flavors of
equal mass $m$ in the Nambu--Jona-Lasinio model.
As a first step and for simplicity, we shall
not study the domain of low temperature and high baryon chemical
potential where the ground state corresponds to a LOFF crystal or to a
color superconductor. We shall also restrict ourselves to small
isospin chemical potential, $\mu_I<\mu_I^{\rm crit}$.
A more complete analysis will be published elsewhere.

{\it Phase Diagram in the Nambu--Jona-Lasinio Model.} -- A crucial
observation made in \cite{KTV} is that when both $\mu_B\neq0$ and
$\mu_I\neq0$ in QCD, there is no reason to expect that the
quark-antiquark condensates are equal for each flavor. Indeed if
we define $\mu_B=\frac12 (\mu_u+\mu_d)$ and
$\mu_I=\frac12(\mu_u-\mu_d)$, the QCD Lagrangian can be written as
\begin{eqnarray}
    \label{LQCD}
{\cal L}_{\rm QCD}=\sum_{f=u,d} \bar{\psi}_f  \left( i \Dirac
- m + \mu_f    \gamma_0 \right) \psi_f  ,
\end{eqnarray}
and there is no symmetry in the QCD Lagrangian that
constrains $\langle\bar{u}u\rangle$ to be equal to
$\langle\bar{d}d\rangle$. Therefore, we have to consider the
quark-antiquark condensates for each flavor separately. In this letter,
we shall concentrate on three observables: The quark-antiquark
condensates for each flavor $\sigma_u=\langle\bar{u}u\rangle$ and
$\sigma_d=\langle\bar{d}d\rangle$, as well as the pion condensate
$\rho=\frac12 \langle \bar{u}\gamma_5d-\bar{d} \gamma_5 u\rangle$.

We use the same Nambu--Jona-Lasinio model with an
instanton-induced four-fermion interaction as Berges and Rajagopal
who studied the QCD phase diagram at nonzero $T$ and $\mu_B$ in
\cite{BR}. After the standard introduction of bosonic fields via a
Hubbard-Stratonovich transformation and integration
over the fermion fields \cite{Klevansky,BR}, we find that
the mean-field free energy is given by
\begin{eqnarray}
  \label{freeEn}
  \Omega&=&\frac1{8G_1}\left(\sigma_u^2+\sigma_d^2+2 \rho^2  \right)
  \nn\\ 
&& \hspace{.1cm} -\Tr \log \left(\begin{array}{cc} h_u& -F^2(\vec{p})
    \rho\gamma_5 \\ F^2(\vec{p})\rho\gamma_5 &  h_d
\end{array} \right),
\end{eqnarray}
where
\begin{eqnarray}
h_f=(i \omega_n+\mu_f) \gamma_0 + i
      \vec{p}\vec{\gamma} +m+F^2(\vec{p}) \sigma_f,
\end{eqnarray}
with $\omega_n=(2n+1)\pi T$,
and the form factor
\begin{eqnarray}
  F(\vec{p})=\Lambda^2/(\vec{p}^2+\Lambda^2)
\end{eqnarray}
is introduced to mimic the effects of asymptotic freedom
\cite{BR}. In the free energy (\ref{freeEn}), we have only kept the
potentially non-vanishing condensates $\sigma_u$, $\sigma_d$, and
$\rho$. We follow Berges and Rajagopal and take the scale
$\Lambda=0.8$ GeV and the coupling constant $G_1=6.47/ \Lambda^2$
which are reasonable phenomenological choices \cite{BR}. They
correspond to $\sigma_u=\sigma_d=0.4$ GeV at
$m$=$T$=$\mu_B$=$\mu_I$=$0$.

In order to proceed we have to compute the excitation energies in
(\ref{freeEn}), and in general solve a fourth order equation in
$\omega_n$. We have not been able to find a short expression for
the general solution. We shall therefore analyze the most relevant
particular cases separately. First at $\mu_B=0$ (i.e.
$\mu_u=-\mu_d$), relying on lattice simulations \cite{KS} , we can
assume that $\sigma_u=\sigma_d=\sigma$. We find that the
excitation energies are given by
\begin{eqnarray}
  \label{excitEnZeroB}
  E_{\pm}=\sqrt{\left(E\pm \mu_I\right)^2+F^4(\vec{p}) \rho^2},
\end{eqnarray}
where $E=\sqrt{\vec{p}^2+(m+F^2(\vec{p})\sigma)^2}$. Thus the
free energy (\ref{freeEn}) becomes
\begin{eqnarray}
  \label{freeEnZeroB}
  \Omega&=&\frac1{4G_1}\left(\sigma^2+\rho^2\right) \nn \\
&& \hspace{.1cm} -\frac6{\pi^2} \int_0^\infty dp \; p^2 \left[ E_\pm+2 T
  \log   \left(1+e^{-E_\pm/T}   \right)   \right].
\end{eqnarray}
This is exactly the same free energy as the one that has been
studied for the phase diagram of QCD with two colors in
\cite{Benoit2}.  It leads to a Bose-Einstein condensation phase
where $\rho\neq0$ when $\mu_I$ is larger than half the pion mass
\cite{Benoit2}. Therefore at $\mu_B=0$, the Nambu--Jona-Lasinio
model agrees with the results obtained in lattice simulations and
in effective theories \cite{KS,SS,TV,KTV}.

We then study $\mu_B\neq0$ and $\mu_I<\mu_I^{\rm crit}$,
which corresponds to the
most relevant phenomenological situations, i.e. when the pion condensate
vanishes.
The free energy (\ref{freeEn}) then separates into a sum over each flavor
\begin{eqnarray}
  \label{freeEnZeroR}
  \Omega&=&\sum_{f=u,d} \Bigg( \frac1{8G_1} \sigma_f^2  \\
&& \hspace{.1cm} -\frac3{\pi^2} \int_0^\infty dp \; p^2 \left[
  E_\pm^f+2 T   \log
  \left(1+e^{-E_\pm^f/T}   \right)   \right] \Bigg), \nn
\end{eqnarray}
where
\begin{eqnarray}
  \label{excitEnZeroR}
E_\pm^f=\sqrt{\vec{p}^2+(m+F^2(\vec{p})\sigma_f)^2}\pm\mu_f.
\end{eqnarray}
The free energy (\ref{freeEnZeroR}) has two remarkable properties.
It is even in $\mu_u$ and $\mu_d$ separately and it can be
expressed as a sum over the different quark flavors. Both of these
properties are also found in the Random Matrix model studied in
\cite{KTV} and are essentially responsible for the striking
changes in the phase diagram.

The evenness of the free energy in $\mu_u$ and $\mu_d$ implies
that the crossover line that separates the hadronic phase from the
quark-gluon-plasma phase in the $\mu_B-T$ plane at $\mu_I=0$
coincides with the corresponding crossover line in the $\mu_I-T$
plane at $\mu_B=0$. This property was indeed found in numerical
lattice simulations \cite{HK,KS}.

Since $\Omega=\sum_{f=u,d} \Omega_f(\mu_f)$, the free energy is
minimized by 
minimizing each $\Omega_f$ separately. But each $\Omega_f$ is equal,
up to a factor two, to the free energy studied by Berges and Rajagopal
for zero diquark condensate in \cite{BR}. Therefore, for
each flavor separately, the domains where $\sigma_f\sim 0.4$ GeV and
$\sigma_f<<0.4$ GeV are separated by a first order line that ends in a
critical endpoint where a crossover lines starts. Now since
$\Omega_u(\mu_u)=\Omega_u(\mu_B+\mu_I)$ and
$\Omega_d(\mu_d)=\Omega_u(\mu_B-\mu_I)$ ,
the whole phase diagram for QCD with two flavors in the $\mu_B-T$
plane at nonzero $\mu_I$
corresponds to a superposition of
two of the usual phase diagrams at $\mu_I=0$ shifted by $2\mu_I$. This is
illustrated in Figure~1 for $m=10$ MeV and for both $\mu_I=0$ and
$\mu_I=30$ MeV. 
\begin{figure*}
\includegraphics[scale=1.0, angle=0, draft=false]{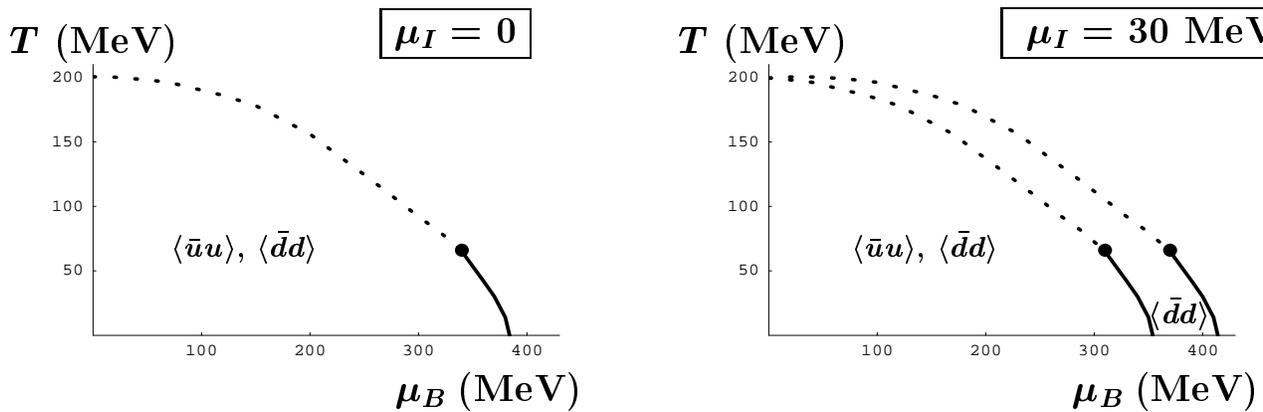}
\caption{Phase diagram in the $\mu_B$-$T$-plane for a quark mass
  $m=10$ MeV and an isospin chemical potential $\mu_I=0$, and
  $\mu_I=30$ MeV, respectively. At low temperature a  first order
  phase transition takes place at the full line that ends in the
  critical endpoint. The dotted curves depict the crossover
  behavior. The condensates that are not displayed are of the order of
  the quark mass. This phase diagram
  can be trusted only above temperatures of a few tens of MeV since
 we did not consider the possibility of a
  crystalline LOFF phase or a color superconductor.
The temperature of the critical endpoint is not affected by the
isospin chemical potential.
\label{fig1}}
\end{figure*}
There are now two first order lines that start at
$T=0$ and which end in a critical endpoint at $T\sim65$ MeV. The
temperature of the critical endpoint is not affected by the isospin
chemical potential. Two crossovers emerge from the critical endpoints
and intersect on the $\mu_B=0$ axis at $T\sim200$ MeV. Notice that the
very low temperature part of this phase diagram cannot be trusted since
we did not consider the possibility of a crystalline LOFF phase or a
color superconductor. Nevertheless the phase diagram above a few tens of
MeV is correct since the LOFF and the color superconducting phases
disappear for such temperatures.

{\it Conclusions and Discussion.} --
In this letter we have used a Nambu--Jona-Lasinio model to show that
the introduction of an isospin
chemical potential leads to qualitative changes in the QCD phase
diagram at nonzero temperature and baryon chemical potential.

First, in agreement with lattice simulations \cite{HK,KS} and the
Random Matrix model analyzed in \cite{KTV}, we find that the
crossover line at low $\mu_B$ and $\mu_I=0$ is identical to the
crossover line at low $\mu_I$ and $\mu_B=0$.

Second, at low temperature, there are two first order phase
transitions that end in two critical endpoints, and there are two
crossovers at low baryon chemical potential. All these results are
in complete agreement with the Random Matrix model studied in
\cite{KTV}. The new lattice techniques developed to study the
small $\mu_B$ behavior at $\mu_I=0$ can be used to study this pair
of crossover lines.

The existence of this pair of crossover lines is important for heavy
ion collision physics. First, as was shown in
\cite{RHICtricPt,RHICtricPt2,Fukushima},
the critical
endpoints have definite signatures that can be observed in these
experiments. The effect of the isospin chemical potential is twofold:
It doubles the number of critical endpoints and pushes one of them to
lower $\mu_B$. The second effect makes the presence of the critical
endpoint easier to see in heavy ion collision experiments.
Second,  the transition from the quark-gluon-plasma to the
hadronic phase will be softer at nonzero $\mu_I$ than at $\mu_I=0$,
since the system has to go through two crossover lines instead of
one.

Finally if the strange quark is included, the hadronic phase and
the quark-gluon-plasma phase are expected to be separated either
by a crossover or by a first order phase transition at zero
$\mu_I$, depending on the precise value of the strange quark mass
\cite{PKLS} . The effect of a small strange quark mass is to push
the critical endpoint towards higher $T$. We expect that the
introduction of a nonzero $\mu_I$ at a physical value of the
strange quark mass will generate two of these separation lines:
Either two crossovers or two first order phase transitions will be
present at high temperature and low baryon chemical potential.

\vspace*{1cm}

{\it Acknowledgments.} -- B. Klein and J. Verbaarschot are
acknowledged for useful discussions. D. T. is supported in part by
the "Holderbank"-Stiftung. This work was partially supported by
the NSF under grant NSF-PHY-0102409.

\end{document}